\documentclass[12pt,a4paper]{article}
\usepackage[T2A]{fontenc}
\usepackage[utf8]{inputenc}
\usepackage[english]{babel}
\usepackage{amssymb}
\usepackage{amsmath}
\usepackage{graphicx}
\usepackage[small,sc,center]{titlesec}
\usepackage{indentfirst}
\usepackage{siunitx}
\usepackage[labelsep=period]{caption}
\usepackage{caption}

\newcommand{\nc}{\newcommand}
\nc{\Teff}{T_\mathrm{eff}}
\nc{\tev}{t_\mathrm{ev}}
\begin{document}

\begin{center}
	\textbf{Evolutionary and hydrodynamic models of short--period Cepheids}
	
	\vskip 3mm
	\textbf{Yu. A. Fadeyev\footnote{E--mail: fadeyev@inasan.ru}}
	
	\textit{Institute of Astronomy, Russian Academy of Sciences,
		Pyatnitskaya ul. 48, Moscow, 119017 Russia} \\
	
	Received April 16, 2020; revised April 26, 2020; accepted April 28, 2020
\end{center}

\textbf{Abstract} ---
The evolutionary calculations for population I stars with masses on the main sequence
$5 M_\odot\le M_0 \le 6.1 M_\odot$ and initial fractional abundances of helium
$Y_0=0.28$ and heavier elements $Z_0=0.02$ were carried out to the stage of central
helium exhaustion.
Selected core helium--burning models were used as initial conditions for solution
of the equations of hydrodynamics and time--dependent convection describing radial
pulsations of Cepheids.
In the Hertzsprung--Russel diagram the evolutionary tracks are shown to  cross the red
edge of the instability strip for $M_0 > 5.1M_\odot$.
The grid of hydrodynamic models of core helium--burining Cepheids on the stages of
the second and the third crossings of the instability strip was computed.
The pulsation period $\Pi$ and the rate of period change $\dot\Pi$ were determined as
a continuous function of star age for each evolutionary sequence of
first--overtone pulsators.
Results of calculations agree with observational estimates of $\dot\Pi$ recently
obtained for the short--period Cepheids V532~Cyg, BG~Cru and RT~Aur.


\newpage
\section*{introduction}

Pulsating variable stars of Population~I with nearly symmetrical light curves of the small
amplitude (($\Delta V\le 0.5$ mag) and with periods shorter than 7 day are classified as
short--period Cepheids.
The General Catalogue of Variable Stars (Samus et al. 2017) lists as many as 50 stars
satisfying this criterion where they are designated as DCEPS type variables.
The short--period Cepheids belong to the group of least luminous and least massive Population~I
Cepheids.
Most of the DCEPS variables are assumed to be the first overtone pulsators.

The intermediate--mass stars become the Cepheid variables during the core helium--burning
stage when the star leaves the red giant domain and evolves in the Hertzsprung--Russel (HR)
diagram  along the loop crossing the pulsation instability strip
(Hofmeister et al. 1964; Iben 1966).
Contraction and subsequent expansion of the star during this stage of evolution is due to
changes of the mean molecular weight and opacity of the stellar matter in the envelope
surrounding the convective core (Walmswell et al. 2015).
Extention of the loop in the HR diagram decreases with decreasing stellar mass and is
less sensitive to variations of the Cepheid chemical composition
(Pietrinferni et al. 2006; Bertelli et al. 2009).
Ambiguity in estimates of the Cepheid lower mass limit is due to uncertainties arising from
comparison of the theoretically computed bolometric luminosity $L$ and effective temperature
$\Teff$ with empirical instability strip edges expressed in terms of the absolute magnitude
$M_V$ and the intrinsic color $(B-V)_0$ (Tammann et al 2003).

The history of systematic light measurements of many Cepheids is as long as 120 yr.
For stars with periods shorter than 7 day this interval encompasses more than
$6\times 10^3$ pulsation cycles, so that the rates of secular period change $\dot\Pi$
can be reliably determined from analysis of $O-C$ diagrams (Turner et al. 2006).
The goal of the present study is to determine the fundamental parameters of
short--period Cepheids with recent observational estimates of the period change rate $\dot\Pi$.
To this end we employ the method based on consistent calculations of stellar evolution and
nonlinear stellar pulsations.
Earlier this method was employed for determination of the fundamental parameters of
long--period Cepheids (Fadeyev 1018b).
In this study of evolutionary and hydrodynamic models of short--period Cepheids
we also try to evaluate the lower mass limit of Cepheids.

\section*{methods of computation}

Evolutionary sequences of the Cepheid models were computed using the program MESA version
12115 (Paxton et al. 2018).
The nucleosynthesis kinetic equations were solved for the reaction network
'pp\_cno\_extras\_o18\_ne22.net'
consisting of 26 isotopes from hydrogen ${}^1\mathrm{H}$ to magnesium ${}^{24}\mathrm{Mg}$
coupled by 81 reactions.
The rates of nuclear reactions were computed using the data base JINA Reaclib 
(Cyburt et al. 2010).
The mass loss rate $\dot M$ was calculated according to Reimers (1975) with parameter
$\eta_\mathrm{R} = 0.3$.
Convective mixing of the stellar matter was treated in the framework of the standard theory
by B\"ohm--Vitense (1958) with mixing length to pressure scale height ratio
$\alpha_\mathrm{MLT} = \ell/H_\mathrm{P} = 1.6$.
Extra mixing due to overshooting beyond the boundaries of convective zones was computed
following the approach developed by Herwig (2000):
\begin{equation}
 D_\mathrm{ov}(z) = D_0 \exp\left(-\frac{2z}{f H_\mathrm{P}}\right) ,
\end{equation}
where $D_0$ is the diffusion coefficient (Langer et al. 1985) inside the convection zone
on the distance $0.004 H_\mathrm{P}$ from the convection zone boundary,
$z$ is the spatial coordinate measured from the boundary of convective instability,
$f=0.016$ is the overshooting parameter.

Thermonuclear helium burning in the intermediate--mass stars occurs in the convective core
so that the helium abundance jump appears on the upper boundary of the convection zone.
During stellar evolution the amplitude of this jump increases due to gradual decrease of
the central helium abundance
The core growth and displacement of the upper boundary of the convection zone along
the Lagrangian mass zones is accompanied by irregular ingestion of helium--rich material
which is responsible for sharp increases of nuclear energy production and appearence
of spurious loops on the evolutionary tracks.
In general the effect of core breathing pulses substantially prolongates the core
helium--burning evolutionary stage (Constantino et al. 2016) and leads to significant mistakes
in theoretical estimates of the Cepheid period change rates.
In the present study to avoid the abrupt changes of central helium abundance during
the core helium--burning stage we employed the method which restricts the mass flux on the
outer boundary of the convective core (Spruit 2015; Constantino et al. 2017).
Evolutionary calculations of the core helium--burning stage were carried out with
the time step limit $\Delta t \le 10^2$ yr and the number of Lagrangian mass zones
of the stellar model was $\approx 1.5\times 10^4$.

Selected models of evolutionary sequences corresponding to the core helium--burning Cepheid
stage were used as initial conditions in solution of the equations of radiation
hydrodynamics and time--dependent convection describing radial stellar oscillations.
The inner boundary of hydrodynamic models was considered as the rigid permanently emitting
sphere of radius $r_0=0.1R$, where $R$ is the stellar radius.
Initial values of mesh variables required for the solution of the Cauchy problem were determined
by nonlinear interpolation of the evolutionary model mesh variables.
The method of the solution of the Cauchy problem is described in our earlier paper (Fadeyev 2015).
Hydrodynamic models were computed on the nonuniform Lagrangian mesh consisting of $N=500$
mass zones.
The Lagrangian mass intervals increase inward as the geometrical progression with the
constant factor $q\approx 1.03$.
The oscillation period $\Pi$ of each hydrodynamic model was determined using the
discrete Fourier transform of the kinetic energy of pulsation motions $E_\mathrm{K}$
(Fadeyev 2018a).

\section*{results of computation}

In the present study the initial conditions were presented by 12 evolutionary sequences of
stars with initial masses $5M_\odot\le M_0\le 6.1M_\odot$ that were computed with the
initial mass step $\Delta M_0 = 0.1M_\odot$.
The initial helium and metal fractional abundances on the main sequence ($\tev=0$) were assumed
to be $Y_0=0.28$ and $Z_0=0.02$, respectively.
Each evolutionary sequence was computed to helium exhaustion in the stellar core
($Y_\mathrm{c}\le 10^{-4}$).

\begin{figure}
\centerline{\includegraphics{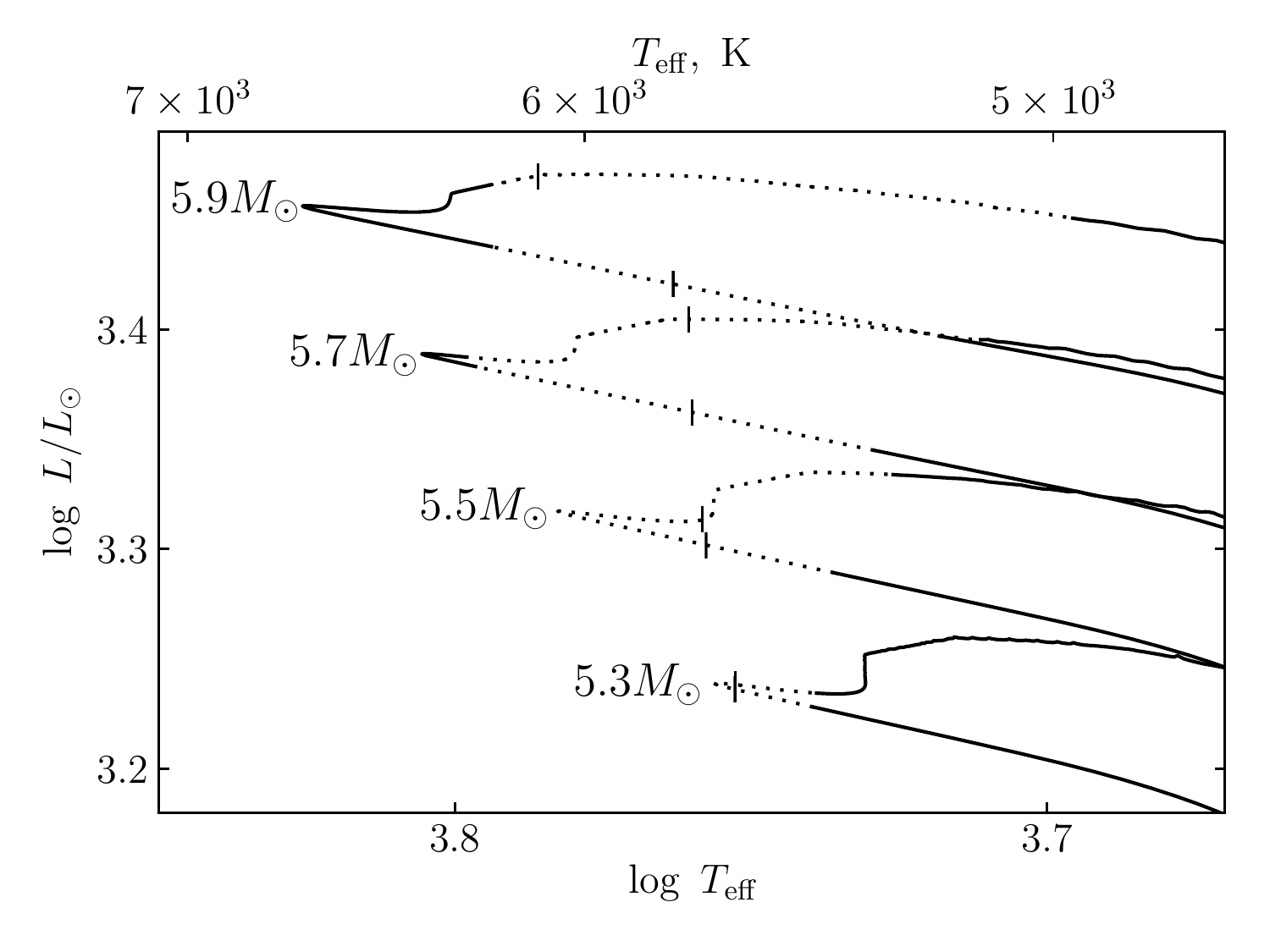}}
\caption{Evolutionary tracks in the HR diagram in vicinity of the Cepheid instability strip.
         Dotted lines show the parts of the track corresponding to instability against
         radial oscillations.
         Vertical dashes on the tracks indicate the point of mode switching between the fundamental mode
         and the first overtone.
         Numbers at the curves indicate the initial mass $M_0$.}
\label{fig1}
\end{figure}

Results of consistent stellar evolution and nonlinear stellar pulsation calculations are
illustrated in Fig.~\ref{fig1}, where evolutionary tracks of stars with initial masses
$M_0=5.3M_\odot$, $5.5M_\odot$, $5.7M_\odot$ and $5.9M_\odot$
are shown in vicinity of the Cepheid instability strip of the HR diagram.
Dotted lines indicate the stage of evolution corresponding to positive growth rate of
the pulsation kinetic energy ($\eta = \Pi^{-1} d \ln E_{\mathrm{K}\max}/dt > 0$).
Here $t$ is time connected with stellar pulsations,
$E_{\mathrm{K}\max}$ is the maximum of the kinetic energy of pulsation motions.
Radial oscillations of Cepheids are described with good accuracy in terms of
standing waves, so that the kinetic energy $E_\mathrm{K}$ reaches its maximum
$E_{\mathrm{K}\max}$ twice per period $\Pi$.

The age of the star $t_{\mathrm{ev},0}$ on the instability strip edge ($\eta=0$)
was determined by linear interpolation between two adjacent models with
opposite signs of the growth rate $\eta$ (Fadeyev 2013).
For considered evolutionary sequences the lower mass limit of Cepheids is in the range
$5.1M_\odot < M_0 < 5.2M_\odot$.
In particular, the turning point of the evolutionary track with $M_0 = 5.2M_\odot$
corresponds to the effective temperature $\log\Teff=3.742$ which
exceeds the effective temperature of the red edge by $\Delta\log\Teff=3.4\times 10^{-3}$.
At the turning point the star is the fundamental mode pulsator with period
$\Pi \approx 5.3$~day.

Evolution of the core helium--burning stars shown in Fig.~\ref{fig1} proceeds clockwise
along the loop where the vertical dashes indicate the oscillation mode switch.
As in our earlier paper (Fadeyev 2019) in this study we assume that the time of mode switch
is much less in comparison with nuclear evolution time of the Cepheid.
The star age $t_\mathrm{ev,sw}$ at the mode switch was evaluated as a mean age of two
adjacent models pulsating in different modes.
Oscillations in the fundamental mode take place in stars with lower effective temperatures
near the red edge of the instability strip whereas oscillations in the first overtone
occur in stars with higher effective temperatures located in the HR diagram near the blue
edge of the instability strip.

\begin{figure}
\centerline{\includegraphics{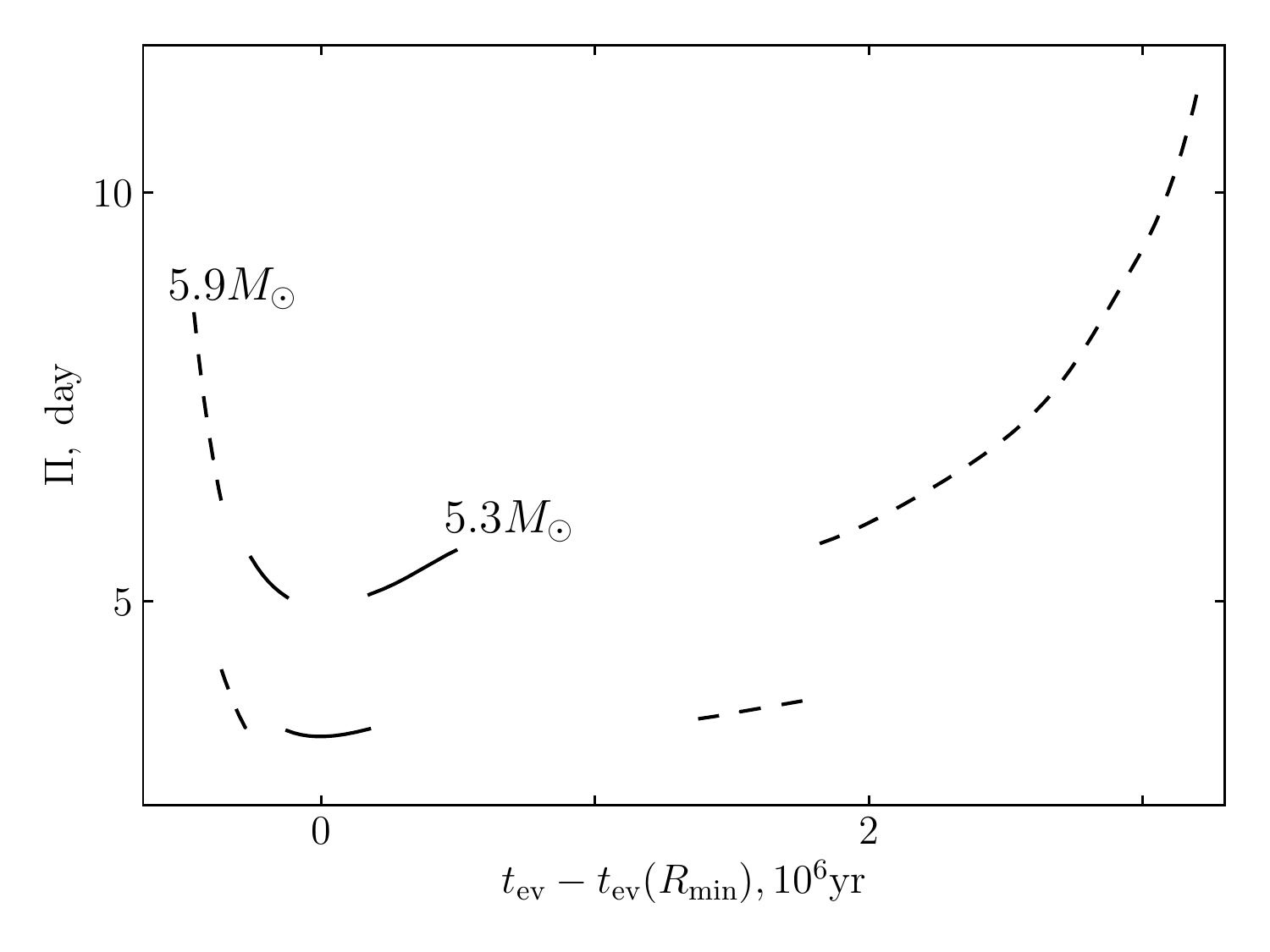}}
\caption{Evolutionary variations of the period of radial pulsations of Cepheids with initial masses
         $M_0=5.3M_\odot$ (solid lines) and $M_0=5.9M_\odot$ (dashed lines).
         Here $\tev(R_\mathrm{min})$ is the star age at the turning point of the evolutionary track
         where the stellar radius reaches its minimum.}
\label{fig2}
\end{figure}

The pulsation period as a continuous function of evolutionary time was determined using the
cubic interpolating splines within the whole interval of $\tev$ where the star is either
the fundamental mode or the first overtone pulsator.
The plots of evolutionary changes of the pulsation period in Cepheid models with initial masses
$M_0=5.3M_\odot$ and $5.9M_\odot$ are shown in Fig.~\ref{fig2}, where for the sake of
graphical representation the horizontal coordinate is set to zero at the turning point of
the evolutionary track where the stellar radius reaches it minimum.

The initial and the final points of each plot in Fig.~\ref{fig2} represent the cross
of the red edge of the instability strip by the evolutionary track, whereas the abrupt
change of the period by a factor of $\approx 1.5$ is due to pulsation mode switch.
The gap in the dependence $\Pi(\tev)$ for the first overtone oscillations of the evolutionary
sequence $M_0=5.9M_\odot$ is due to the fact that the turning point of the evolutionary track
is beyond the blue edge of the instability strip.
As is seen, increase of the Cepheid mass is accompanied by increasing ratio
$\Delta t_{\mathrm{ev},3}/\Delta t_{\mathrm{ev},2}$,
where $\Delta t_{\mathrm{ev},2}$ and $\Delta t_{\mathrm{ev},3}$
are Cepheid lifetimes during the second and the third crossings of the instability strip.

For evolutionary sequences considered in the present study the phenomenon of mode switch
between the fundamental mode and the first overtone occurs in models with radius
$R\le 52R_\odot$.
In particular, in evolutionary sequences with initial masses $5.3\le M_0\le 5.9M_\odot$
mode switching occurs during both the second and the third crossings of the instability strip,
whereas for $6.0M_\odot\le M_0\le 6.1M_\odot$ mode switching occurs only during
the second crossing since on the stage of the third crossing the stellar radius is greater than
the threshold radius: $R > 52R_\odot$.

The region of excitation of stellar pulsations in Cepheids locates in the layers of partial
helium ionization so that the necessary condition for the first overtone pulsation
is that the radius of the helium ionizing zone remains larger than the radius of the node
of the corresponding eigenfunction.
Unfortunately, the general condition for mode switching cannot be defined by simple
relationships.
Below we will use the mean density of stellar matter
$\langle\rho\rangle = M/(\frac{4}{3}\pi R^3)$
as a parameter connected with the thershold period values of the fundamental mode
$\Pi_0$ and the first overtone $\Pi_1$ corresponding to mode switching.

\begin{figure}
\centerline{\includegraphics{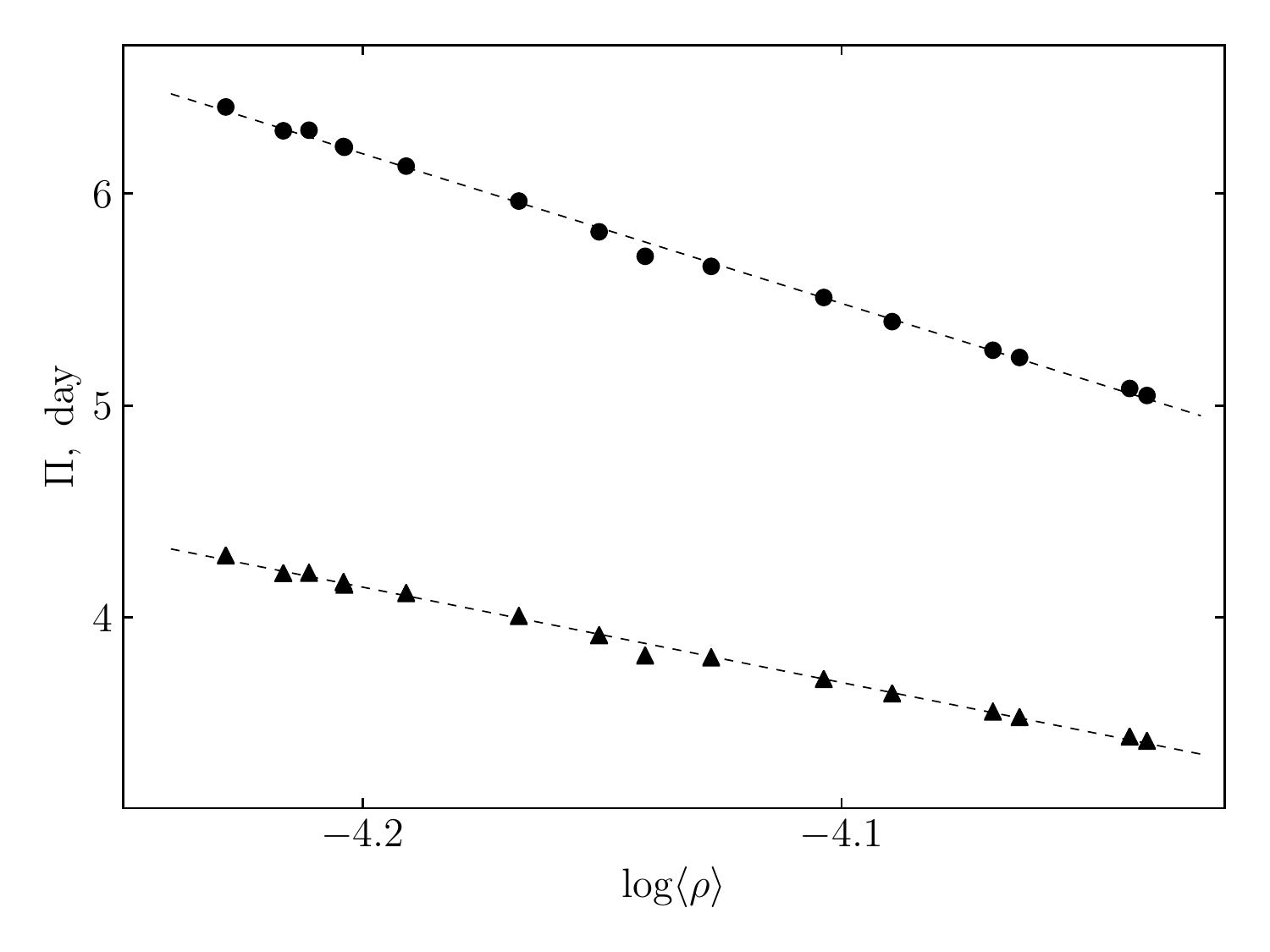}}
\caption{Threshold periods of the fundamental mode (filled circles) and the first overtone (filled triangles)
         versus mean density $\langle\rho\rangle$.
         Relations (\ref{p0msw}) and (\ref{p1msw}) are shown by dashed lines.}
\label{fig3}
\end{figure}

The thershold values $\Pi_0$ and $\Pi_1$ were determinated for all evolutionary sequences
and are described with good accuracy by relations
\begin{equation}
\label{p0msw}
\Pi_0 = - 23.486 - 7.065 \log\langle\rho\rangle ,
\end{equation}
\begin{equation}
\label{p1msw}
\Pi_1 = - 14.773 - 4.504 \log\langle\rho\rangle ,
\end{equation}
where the pulsation period is expressed in days.
Plots of $\Pi_0$ and $\Pi_1$ versus mean density $\langle\rho\rangle$
as well as relations (\ref{p0msw}) and (\ref{p1msw}) are shown in Fig.~\ref{fig3}.

\section*{comparison with observations}

\begin{figure}
\centerline{\includegraphics{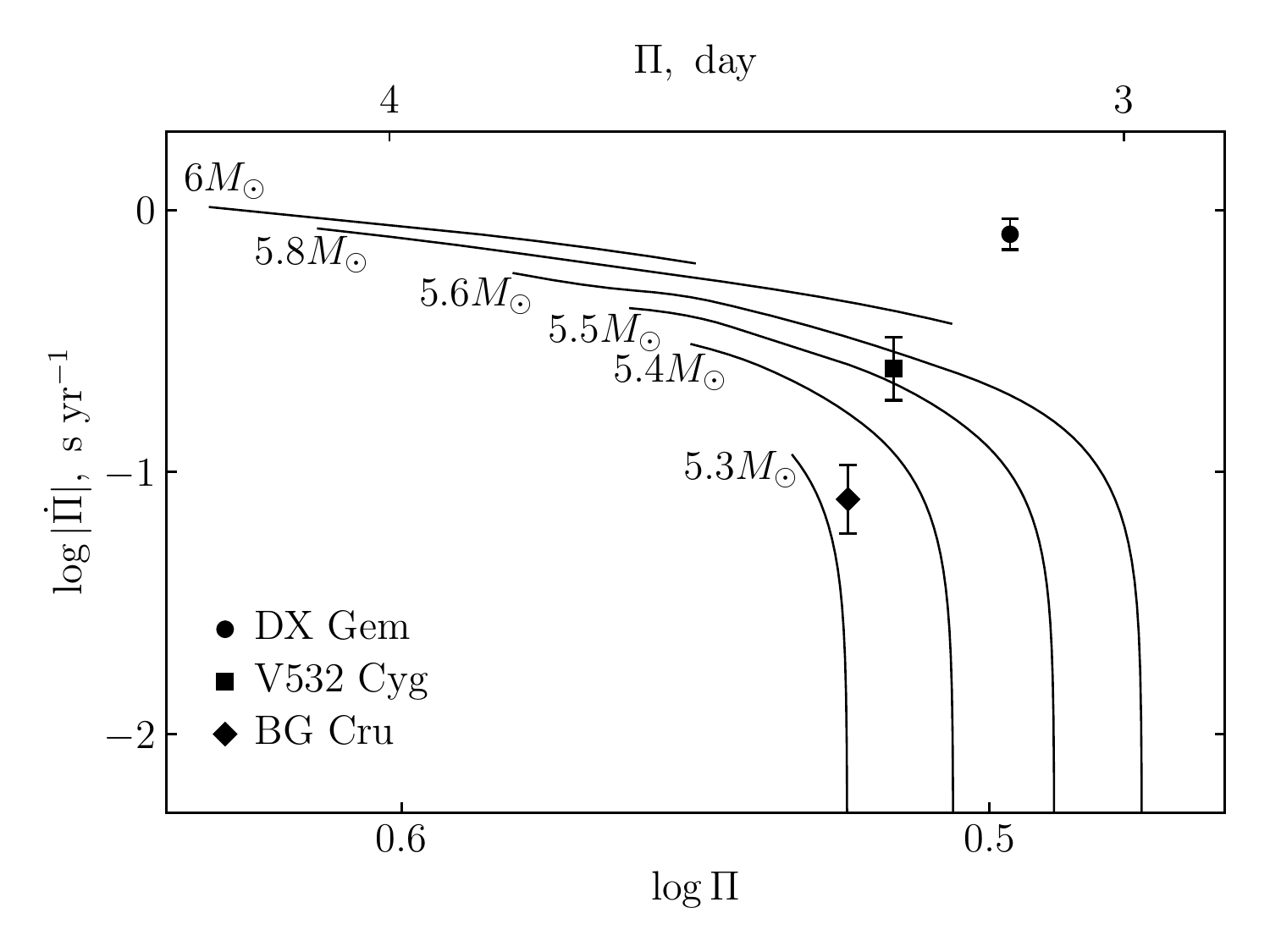}}
\caption{Period change rate $\dot\Pi$ versus period $\Pi$ for Cepheid models
         pulsating in the first overtone during the second crossing of the
         instability strip ($\dot\Pi < 0$).}
\label{fig4}
\end{figure}

The diagram period--period change rate provides the best way to compare
the results of theoretical calculations with observational estimates
of $\Pi$ and $\dot\Pi$ (Turner et al 2006; Fadeyev 2014).
Approximation of $\Pi(\tev)$ by cubic interpolation splines allows us
to determine the period change rate as a continuous function of evolutionary time
and finaly to express $\dot\Pi$ as a function of $\Pi$.
Plots of $\dot\Pi$ versus $\Pi$ for the Cepheid models pulsating in the first
overtone during the second crossing of the instability strip ($\dot\Pi < 0$)
are shown in Fig.~\ref{fig4}.
Stellar evolution proceeds from left to right with decreasing period.
Each plot represents variation of $\dot\Pi$ from the point of mode switching to
the blue edge of the instability strip ($M_0 \ge 5.8M_\odot$) or to
the turning point of the evolutionary track ($M_0 \le 5.6M_\odot$) where $\dot\Pi=0$.

Also shown in Fig.~\ref{fig4} are the recent observational estimates of $\Pi$ and $\dot\Pi$
for short--period Cepheids DX~Gem (Berdnikov 2019a), V532~Cyg (Berdnikov 2019b) and BG~Cru
(Berdnikov et al. 2019).
As is seen, the results of computations agree with observations of
V532~Cyg and BG~Cru, whereas the observational estimate of $\dot\Pi$ in DX~Gem
is several times higher than it can be predicted from our evolutionary sequences.
In order to avoid the discrepancy one should assume that the blue edge of the instability
strip of the star with mass $M\approx 6M_\odot$ is displaced to higher effective temperatures
and therefore to shorter pulsation periods.

\begin{figure}
\centerline{\includegraphics{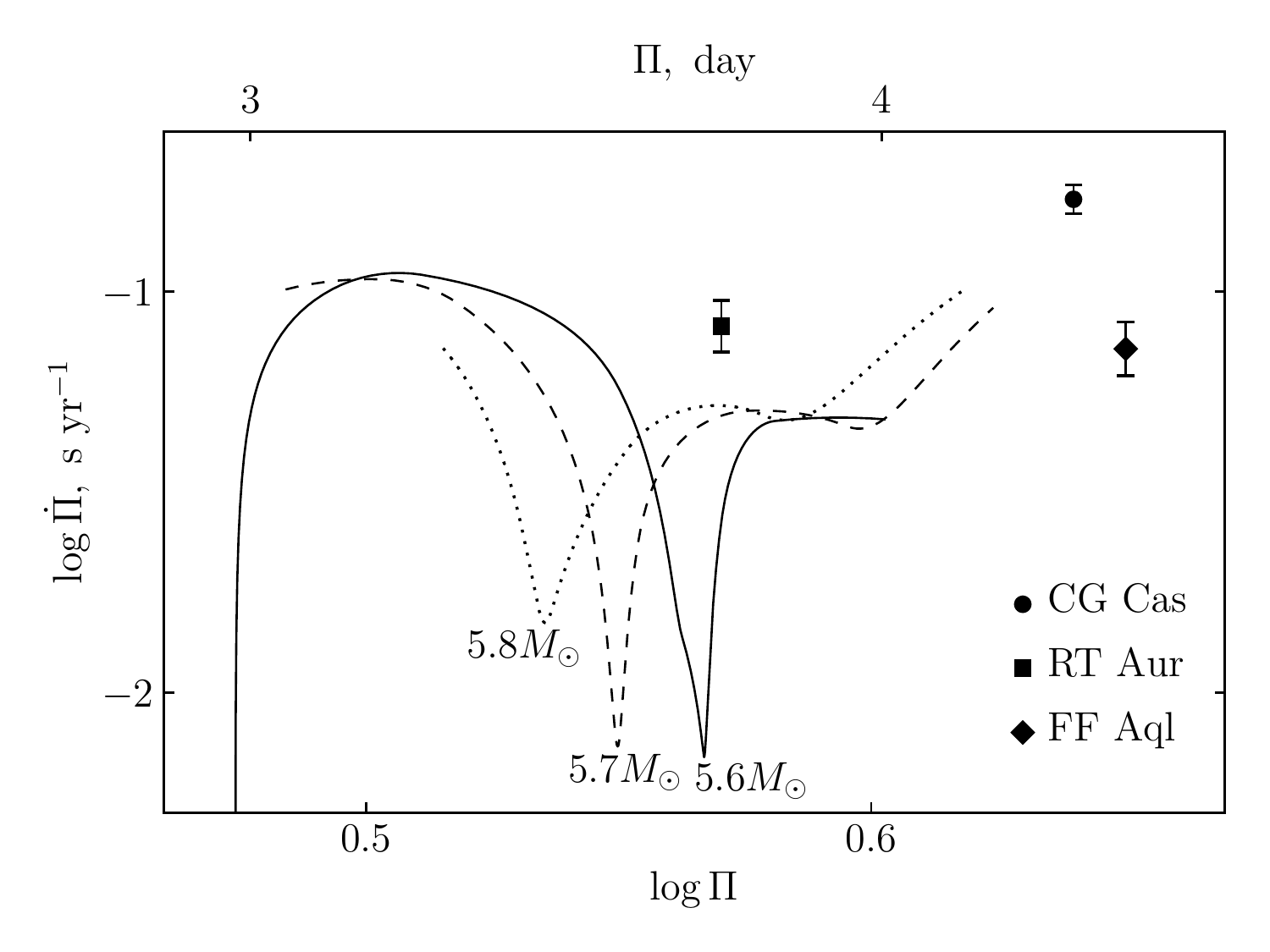}}
\caption{Same as Fig.~\ref{fig4} but for Cepheid models during the third crossing
         of the instability strip ($\dot\Pi > 0$).}
\label{fig5}
\end{figure}

Plots of $\dot\Pi$ versus $\Pi$ for the third crossing of the instability strip
($\dot\Pi >0$) are shown in Fig.~\ref{fig5} for evolutionary sequences
$M_0=5.6M_\odot$, $5.7M_\odot$ and $5.8M_\odot$.
As is seen, all the plots are in satisfactory agreement with RT~Aur (Turner et al. 2007)
since its observational estimate of $\dot\Pi$ is only one and a half times as high as
results of our calculations.
Fig.~\ref{fig5} shows also the observational estimates of $\Pi$ and $\dot\Pi$ for
Cepheids CG~Cas (Turner et al. 2008) and FF~Aql (Berdnikov et al. 2014).
Unfortunately, periods of these Cepheids significantly exceed the thershold values $\Pi_1$
determined in the present study.

In Table~\ref{tabl1} we list the fundamental parameters of the short--period Cepheids
V532~Cyg, BG~Cru and RT~Aur for which the theoretical estimates of $\dot\Pi$ are in
agreement with observations.

\section*{conclusions}

In this paper we extended the results of our earlier study (Fadeyev 2014) devoted to
the theoretical explanation of the observed secular period changes in Cepheids and where
the short--period Cepheids were not considered in detail.
In particular, results of the present work demonstrate that siginificant scatter of the
observational estimates of $\dot\Pi$ in short--period Cepheids is due to the fact
that at the turning point of the evolutionary track the period change rate reduces to zero.
Therefore, the near--zero value of $\dot\Pi$ allows us to roughly evaluate
the Cepheid mass.
As is seen in Fig.~\ref{fig1}, the turn of the evolutionary track inside
the instability strip occurs in stars with masses $M < 5.7M_\odot$.

Results of consistent stellar evolution and nonlinear stellar pulsation calculations
allow us to conclude that the lower mass limit of Cepheids on the stage of core
helium--burning is $M\approx 5.1M_\odot$.
Evolutionary tracks of less massive stars do not cross the red edge of the
instability strip and the star remains stable against radial oscillations.
One should however to note that this conclusion is based on the results
obtained for initial fractional abandances of helium and heavier elements
$Y_0=0.28$ and $Z_0=0.02$.
Sensitivity of evolutionary changes of the stellar radius and of the
extension of the loop of the evolutionary track to the mean molecular
weight (Walmswell et al. 2015) allows us to suppose that the Cepheid lower
mass limit depends on the chemical composition of stellar matter.

Among six Cepheids with observational estimates of $\dot\Pi$ considered
in the present study we succeeded to obtain a good agreement with
observations only for three of them.
The periods of CG~Cas ($\Pi=4.3656$~day) and FF~Aql ($\Pi=4.4709$~day)
are noticeably greater than the threshold period of the first overtone
$\Pi_1$ corresponding to switch to the fundamental mode (see Fig.~\ref{fig5}).
Dependence of the mode switching threshold periods on the mean density
(see Fig.~\ref{fig3}) implies that the upper first overtone period limit
might noticeably depend on the chemical composition of stellar matter.
Therefore, to reconcile the theory with observations of short--period Cepheids
CG~Cas and FF~Aql we have to consider more extended grids of evolutionary and
hydrodynamic models computed for various initial compositions.

\newpage
\section*{references}

\begin{enumerate}
\item L.N. Berdnikov, Astron. Lett. \textbf{45}, 435 (2019a).

\item L.N. Berdnikov, Astron. Lett. \textbf{45}, 677 (2019b).

\item L.N. Berdnikov, D.G. Turner, and A.A. Henden, Astron. Rep. \textbf{58}, 240 (2014).

\item L.N. Berdnikov, A.Yu. Kniazev, V.V. Kovtyukh, V.V. Kravtsov, T.V. Mishenina,
      E.N. Pastukhova, and I.A. Usenko,
      Astron. Lett. \textbf{45}, 445 (2019).

\item G. Bertelli, E. Nasi, L. Girardi, and P. Marigo,
      Astron. Astrophys. \textbf{508}, 355 (2009).

\item E. B\"ohm--Vitense, Zeitschrift f\"ur Astrophys. \textbf{46}, 108 (1958).

\item T. Constantino, S.W. Campbell, W. Simon, J,C. Lattanzio, and A. van Duijneveldt,
      MNRAS, \textbf{456}, 3866 (2016).

\item T. Constantino, S.W. Campbell, and J.C. Lattanzio, MNRAS \textbf{472}, 4900 (2017).

\item R.H. Cyburt, A.M. Amthor, R. Ferguson, Z. Meisel, K. Smith,
      S. Warren, A. Heger, R.D. Hoffman, T. Rauscher, A. Sakharuk, H. Schatz,
      F.K. Thielemann, and M. Wiescher,
      Astrophys. J. Suppl. Ser. \textbf{189}, 240 (2010).

\item Yu.A. Fadeyev, Astron. Lett. \textbf{39}, 746 (2013).

\item Yu.A. Fadeyev, Astron.Lett. \textbf{40}, 301 (2014).

\item Yu.A. Fadeyev, MNRAS \textbf{449}, 1011 (2015).

\item Yu.A. Fadeyev, Astron. Lett. \textbf{44}, 616 (2018a).

\item Yu.A. Fadeyev, Astron. Lett. \textbf{44}, 782 (2018b).

\item Yu.A. Fadeyev, Astron. Lett. \textbf{45}, 353 (2019).

\item F. Herwig, Astron. Astrophys. \textbf{360}, 952 (2000).

\item E. Hofmeister, R. Kippenhahn, and A. Weigert,
      Zeitschrift f\"ur Astrophys. \textbf{60}, 57 (1964).

\item I. Iben, Astrophys.J. \textbf{143}, 483 (1966).

\item N. Langer, M. El Eid and K.J. Fricke, Astron. Astrophys. \textbf{145}, 179 (1985).

\item B. Paxton, J. Schwab,  E.B. Bauer, L. Bildsten, S. Blinnikov,
      P. Duffell, R. Farmer,  J.A. Goldberg, P. Marchant, E. Sorokina, A. Thoul,
      R.H.D. Townsend, and F.X. Timmes,
      Astropys. J. Suppl. Ser. \textbf{234}, 34 (2018).

\item A. Pietrinferni, S. Cassisi, M. Salaris, and F. Castelli,
      Astrophys. J. \textbf{642}, 797 (2006).

\item D. Reimers, \textit{Problems in stellar atmospheres and envelopes}
      (Ed. B. Baschek, W.H. Kegel, G. Traving, New York: Springer-Verlag, 1975), p. 229.

\item N.N. Samus', E.V. Kazarovets, O.V. Durlevich, N.N. Kireeva, and E.N. Pastukhova,
      Astron. Rep. \textbf{61}, 80 (2017).

\item H.C. Spruit, 2015, Astron. Astrophys. \textbf{582}, L2 (2015).

\item G.A. Tammann, A. Sandage, A. and B. Reindl,
      Astron. Astrophys. \textbf{404}, 423 (2003).

\item D.G. Turner, David G., M. Abdel--Sabour Abdel--Latif, and L.N. Berdnikov,
      Publ. Astron, Soc. Pacific \textbf{118}, 410 (2006).

\item D.G. Turner, I.S. Bryukhanov, I.I. Balyuk, A.M. Gain, R.A. Grabovsky, V.D. Grigorenko,
      I.V. Klochko, A. Kosa--Kiss, A.S. Kosinsky, I.J. Kushmar, V.T. Mamedov,
      N.A. Narkevich, A.J. Pogosyants, A.S. Semenyuta, I.M. Sergey, V.V. Schukin,
      J.B. Strigelsky, V.G. Tamello, D.J. Lane, and D.J. Majaess,
      Publ. Astron, Soc. Pacific \textbf{119}, 1247 (2007).

\item D.G. Turner, D. Forbes, D. English, P.J.T. Leonard, J.N. Scrimger,
      A.W. Wehlau, R.L. Phelps, L.N. Berdnikov, and E.N. Pastukhova,
      MNRAS \textbf{388}, 444 (2008).

\item J.J. Walmswell, J. C.A. Tout, and J.J. Eldridge, MNRAS \textbf{447}, 2951 (2015).

\end{enumerate}

\newpage
\begin{table}
\caption{Fundamental parameters of short--period Cepheids}
\label{tabl1}
\begin{center}
 \begin{tabular}{lcccrrr}
  \hline
   & $\Pi$, day & $\tev,\ 10^6$ yr &  $M/M_\odot$ & $L/L_\odot$ & $R/R_\odot$ & $\Teff$, K \\
  \hline
V532 Cyg & 3.2836  & 82.3 & 5.53 & 2130 & 42.8 & 6000 \\
BG Cru   & 3.3426  & 90.0 & 5.33 & 1810 & 42.8 & 5760 \\
RT Aur   & 3.7182  & 76.1 & 5.77 & 2720 & 47.9 & 6030 \\
  \hline          
 \end{tabular}
\end{center}
\end{table}
\clearpage

\newpage
\clearpage

\newpage
\clearpage

\newpage
\clearpage

\newpage
\clearpage

\newpage
\clearpage

\end{document}